# Was there a negative vacuum energy in your past?

George Chapline[1] and James Barbieri[2]

[1]Lawrence Livermore National Laboratory, Livermore, CA
[2]NAWC-WD, China Lake, CA

A model for gravitational collapse where the event horizon is a quantum critical phase transition is extended to provide an explanation for the origin of the observable universe, where the expanding universe that we observe today was proceeded by a flat universe with a negative cosmological constant. In principal this allows one derive all the features of our universe from a single parameter: the magnitude of the pre-big bang negative vacuum energy density. In this paper a simple model for the big bang is introduced which allows us to relate the present day energy density and temperature fluctuations of the CMB, to the present day density of dark matter. This model for the big bang also makes a dramatic prediction: dark matter mostly consists of compact objects with a masses on the order of $10^4$ solar masses. Remarkably this is consistent with numerical simulations for how primordial fluctuations in the density of dark give rise to the observed inhomogeneous distribution of matter in our universe. Our model for the big bang also allows for the production of some compact objects with masses greater than $10^4$ solar masses, which is consistent with numerical simulations of structure formation which require massive primordial comapact objects as the seeds for galaxies in order to explain galactic morphologies.

One of the outstanding puzzles of modern theoretical physics is that classical general relativity offers no clue as the fate of massive steller cores undergoing gravitational collapse or the state of matter prior to the "big bang". These puzzles are all the more perplexing because In quantum mechanics it is not possible for matter to simple appear or disappear. Prevoiusly we have drawn attention [1,2] to the fact that the quantum critical phase transition theory of event horizons [3] provides a plausible explanation for the fate of matter undergoing gravitational collapse; namely, most of the mass-energy of the collapsing matter is converted into vacuum energy, resulting in the formation of a "dark energy star"[4]. Dark energy stars are distinguished from black holes in that their interiors resemble de Sitter's "interior" solution [5] rather than a black hole space-time. In this paper we offer a possible resolution of the enigma of what proceeded the big bang by noting that a flat Robertson-Walker universe with a negative cosmological constant will naturally evolve to an expanding inhomogeneous universe containing radiation and dark mattter via the same kind of quantum dynamics that resolves the problem of gravitational collapse. It was suggested some time ago by de Sitter, Eddington, and Lemaitre [6] that the observable universe may not have had a singular beginning, but instead may have originated from a finite size seed. Lemaitre suggested that this finite seed was a macroscopic quantum object which he called the "primeval atom". Cosmological models incorporating this idea make use of Lemaitre's examples of Robertson-Walker space-times with a positive cosmological constant [7,8]. In the following we describe a model for the origin of our expanding universe, in which the initial state is not a single quantum object, but an infinite assembly of quantum objects. It has has already been noted that [9] that such a two-phase flat space cosmology provides a simple explanation for many of the observed features of our universe, including the entropy and temperature fluctuations of the cosmic ray microwave background. In this paper we describe how a a flat homogeneous Robertson-Walker universe with a negative cosmological constant can evolve to an expanding universe resembling our own. We then show how the parameters of the standard cosmological model as well as the present day inhomogeous structure of our universe might be derived from a single parameter: the magnitude of the initial negative vacuum energy.

The classical gravitational dynamics of a flat universe with a negative cosmological constant necessarily involves collapse to a density singularity. The acceleration of the cosmological scale factor $R(t)$ in a flat Robertson-Walker universe with a cosmological constant is

$$\ddot{R}^2 = -\frac{4\pi G}{3c^2}(\rho + 3p - 2\rho_\Lambda)R \quad , \tag{1}$$

where $\rho$ is the matter density, $p$ is the matter pressure, and $\rho_\Lambda$ is the vacuum energy density. When the vacuum energy density $\rho_\Lambda$ is negative and the matter is a relativistic gas of particles with an adiabatic index 4/3 Eq. 1 has a simple analytic solution [8]:

$$R(\tau) = R_m \left[\frac{1-\cos 4\alpha\tau}{2}\right]^{1/4} \quad , \tag{2}$$

where $\tau$ is the usual Robertson-Walker universal time. The cosmological constant $\Lambda = -3\alpha^2 = 8\pi G\rho_\Lambda/c^2$. Regardless of its maximum value the scale factor collapses to zero in a time $\tau_c \equiv \pi/4\alpha$. At the time $\tau \equiv \pi/4\alpha$ when the scale factor is a maximum the total energy density $\rho + \rho_\Lambda = 0$. As $\tau$ approaches $\pi/2\alpha$ the energy density, which is dominated by the matter density $\rho$, approaches infinity. In Fig. 1 we show the evolution of the scale factor for an initial scale factor $R_m = 10R_g$, where $R_g = c/\alpha$ is the initial gravitational radius for the matter. We also show the light sphere radius $r_c$ for photons emitted at the initial time $\tau=\pi/4\alpha$. Eq. 2 implies that the conformal radius $r_c/R = \sqrt{2}R_g/R_m F(\cos^{-1}(R/R_m)|\pi/4)$, where F is an incomplete elliptic integral of the first kind. As is evident from Fig.1 photons emitted from any point in the negative cosmological constant universe are trapped, which according to Penrose and Hawking would require collapse to a singularity. On the other hand, in the following we will assume that in reality a negative cosmological constant universe does not collapse to a singularity due to quantum effects.

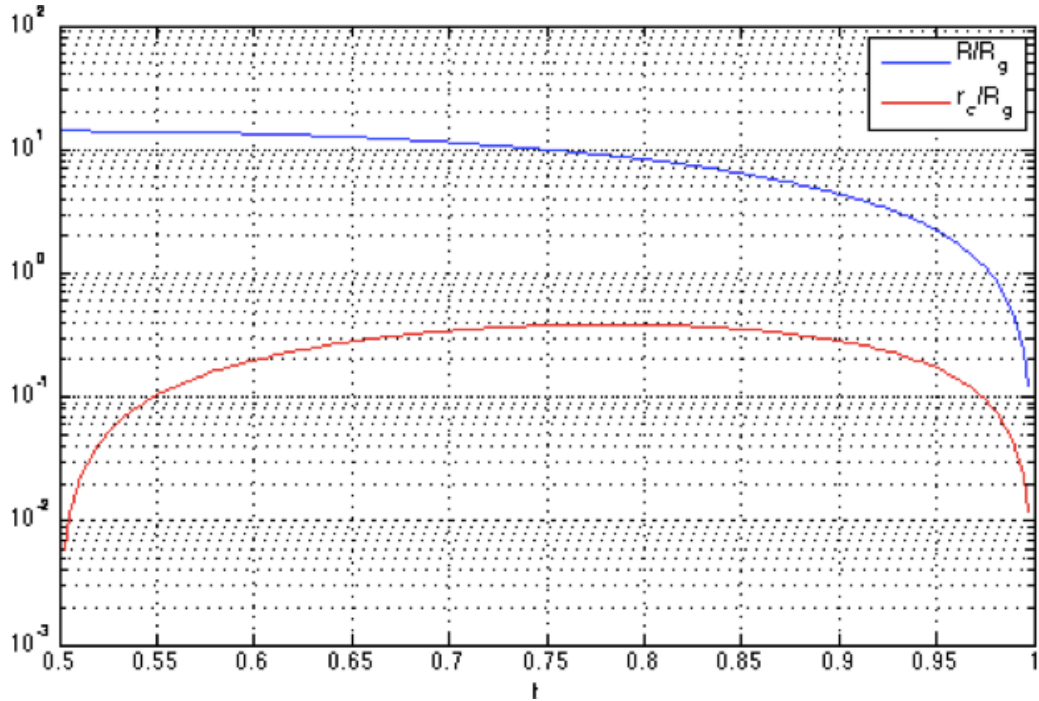

Fig.1. Time evolution of the scale factor in a radiation filled flat Robertson-Walker universe with a negative cosmological constant, together with the light sphere radius for photons emitted at the initial time t ≡ 0.5 Time is measured in units $\pi/2\alpha$, while the radii are measured in units of the initial matter gravitational radius $c/\alpha$.

Our hypothesis is that the same type of conversion of matter mass-energy to vacuum energy [11] that we previously proposed [1-2] as the reason for the avoidance of a singular end point for the gravitational collapse of massive stellar cores will also lead to the avoidance of a mass density singularity in a flat negative cosmological constant universe. In particular we will argue that as a result of the ubiquitous formation of trapped surfaces in a flat negative cosmological constant universe most of the matter mass-energy will be transforned into positive vacuum energy, resulting in an expanding universe which resembles our universe. As a simple model for the conversion of most of the mass-energy of radiation in our negative cosmological constant universe to vacuum energy we propose replacing the usual energy conservation law for a Lemaitre universe with a constant cosmological constant with the equations

$$\frac{d}{dt}(\rho R^3) + p\frac{dR^3}{dt} = -\rho R^3/\tau_c \qquad (3)$$

$$\frac{d\rho_\Lambda}{dt} = \rho/\tau_c \quad , \qquad (4)$$

Numerical solutions of Eq's 1,3. and 4 are shown in Fig.s 2 and 3.

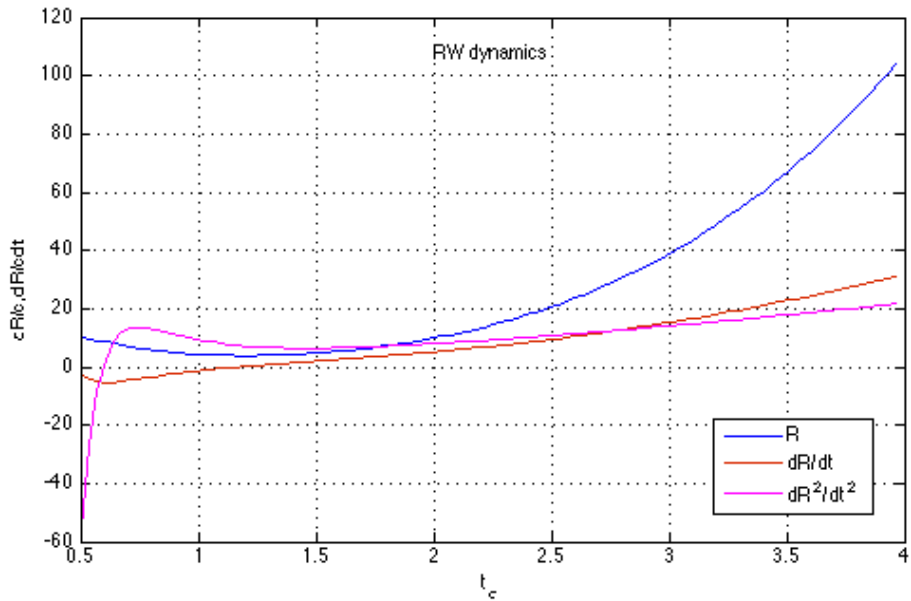

Fig 2 Evolution of a flat Robertson-Walker universe, initially with a negative cosmological constant and filled with radiation, but allowing for the radiation and vacuum energy densities to change according to Eq.s 3 and 4. Time is measured in units $\pi/2\alpha$, while the radii are measured in units of the initial matter gravitational radius $R_g = c/\alpha$.

It can be seen that the acceleration of the Robertson-Walker scale factor switches from being very negative to positive, indicating evolution from a collapsing to an expanding universe. Our model for the big bang consists of Eq.s 1 and 3 together with the stipulation that after the collapse time $\tau_c = \pi/2\alpha$ the vacuum energy created when the de Sitter horizon is small compared to the Hubble radius does not contribute to a cosmological constant, but instead is encapsulated into a form of dark matter.

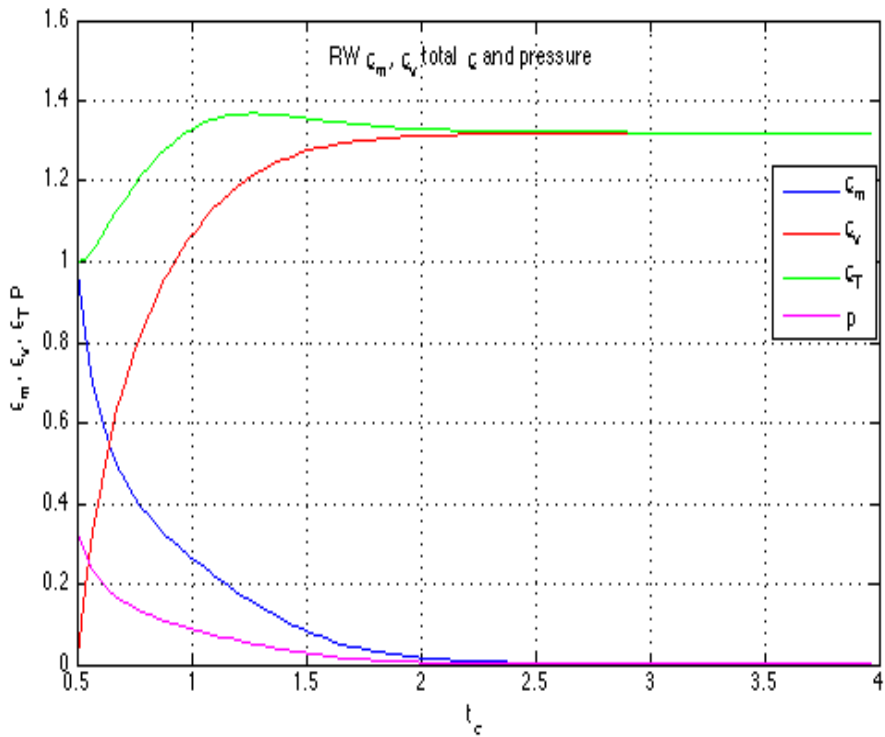

Fig. 3. The matter, vacuum, and total energy densities resulting from the collapse of a flat negative vacuum energy universe

Of course the ultimate fate of matter undergoing gravitational collapse has been a long standing enigma. Following the seminal paper of Oppenheimer and Snyder, it had come to be widely accepted that the gravitational collapse of a sufficiently large mass would inevitably lead to the formation of an event horizon and a density singularity [11]. Moreover, it has generally been believed that these predictions will turn out to be correct even when quantum effects are taken into account, since the formation of an event horizon can take place in a region of space-time where the curvature of space is very small. On the other hand, there are several long standing puzzles connected with the general relativisic picture of gravitational collapse. The most famous of these puzzles concerns the fact that in quantum mechanics information can never disappear. The most likely resolution of this paradox is that quantum effects profoundly affect the classical picture of matter falling smoothly through an event horizon. In particular, there are plausible arguments [3,12,13] that in a quantum theory of gravity the space-time inside an event horizon always resembles de Sitter's "interior" solution of the Einstein equations.

A central element of our argument that a negative cosmological constant evolves into an expanding universe that resembles our own is that, due to the well known instability of infinite de Sitter space at the de Sitter horizon [14], patches space-time resembling de Sitter's interior solution will appear throughout the collapsing universe. These "dark energy stars" are gravitationaly stable, and will initally have a mass

$$M* = 0.3 \ [(GeV)^4/\rho*]^{1/2} \ M_O \ , \qquad (5)$$

where $\rho*$ is the positive vacuum energy created at the collapse time $\tau_c = \pi/2\alpha$ by the conversion of radiation energy in the collapsing negative cosmological universe into vacuum energy and $M_O$ is the mass of the sun. This mass is just the mass inside the de Sitter horizon at the time $\tau_c$.

A two-phase picture for cosmology [15], where space-time is a mixture of ordinary vacuum and dark energy stars, emerges from our model in somewhat the same way that supersaturated steam consists of a mixture of water vapor and water droplets. It is of course rather natural to imagine that in such a picture the initial energy densities of the dark matter and the cosmological vacuum might be comparable. The initial masses of the primordial dark energy stars will be given by Eq.5, but because the spatial density of these dark energy stars will be very large, collisions and fluctuations in the spatial density of the primordial dark energy stars created at $\tau_c \equiv \pi/4\alpha$ will cause them to coalescence (the details are discussed in Ref. 9), leading to formation of more massive compact objects. The reversal of the scale factor acceleration from negative to positive, will result in a universe consisting of dark energy stars and radiation expanding in a Freidmann-like fashion. The maximum mass of these compact objects will be dictated by the time it takes for their spatial density becomes too low for them to continue to coalesce to from larger dark energy stars. We are immediately faced with the puzzle though that expansion of a cloud of dark energy stars with an initial mass $M_*$ would lead to a present day density of matter that is many orders of magnitude larger than the observed dark matter density.

A possible resolution of this puzzle [9] is that when dark energy stars coalesce to form a larger dark energy star the surface area of the resultant dark energy star will be maximized in much the same way that the total black hole surface area increases when two black holes coalesce. Because of this black hole-like behavior a large fraction of the mass-energy of dark energy stars is converted into thermal energy when they coalesce. Our model for the big bang is based on the assumption that this thermal radiation is released as freely streaming radiation when the photon frequency falls below a critical frequency $\nu_c$ where radiation and dark energy stars decouple. The value of this critical frequency was estimated in Ref.s 1,3. If we assume that the gauge field coupling strength at the GUT scale $g^2 = 0.1$, this estimate for the cutoff for strong interactions between dark energy stars and photons is $h\nu_c \approx 1\text{GeV}(M_\oplus/M)^{1/2}$, where $M$ is the mass of the dark energy star and $M_\oplus$ is the mass of the sun.

In our model for the big bang the transition between the very high temperature regime where there is strong coupling between the dark matter and radiation and the lower temperature regime where the dark matter and radiation are decoupled is assumed to be abrupt in the sense that for red shifts greater than a certain red shift, $1+z_r$, the radiation energy is stored as the thermal energy of dark energy stars with masses $M \gg M_*$, while for $1+z < 1+z_r$ we will assume that all the mass-energy of the primordial dark energy stars will have been converted into radiation and remnant dark energy stars with average mass $M_{DM}$. Taking into account the black hole-like relation between the mass and surface area of a dark energy star the cosmological energy density for $1+z < 1+z_r$ will be given by

$$\rho_{DM} = \rho_* \left(\frac{M_*}{M_{DM}}\right)^2 \left(\frac{1+z}{1+z^*}\right)^3, \qquad (6)$$

where $M_{DM}$ is the average mass of a dark matter dark energy star and $1+z_*$ is the red-shift for the break-up of the initial positive vacuum energy state resulting from the collapse of the negative vacuum energy state, corresponding to the origin of the observable universe. As an estimate for the red shift separating these two regimes we will use the value

$$1+z_r = 0.37 h\nu_c / k_B T_{CMB}, \qquad (7)$$

where the factor 0.37 accounts for difference between the temperature and mean photon energy and $T_{CMB}$ = 2.73K is the present day temperature of the CMB. The radiation energy density for $1+z < 1+z_r$ will be given by

$$\rho_{rad} = \rho_* \left(\frac{1+z_*}{1+z_r}\right)\left(\frac{1+z}{1+z_*}\right)^4, \qquad (8)$$

The radiation energy density is related to the radiation temperature $T$ by the usual formula

$$\rho_{rad} = N(T)\frac{\pi^2}{30}\frac{(k_B T)^4}{(\hbar c)^3}, \qquad (9)$$

where $N(z)$ is the effective number of elementary particle species contributing to the radiation energy density at redshift $1+z$. Strictly speaking we should have taken into account $N(T)$ in our estimate, Eq. 7, for the red shift marking the appearance of the CMB, but we have neglected this correction it only depends on $N^{1/4}$.

Combining Eq.s 6-8 with the ratio of the present day energy densities of dark matter (keV/cm³) and the CMB (0.26 eV/cm³) leads to the following relation between $M_{DM}$ and $M_*$ :

$$\left(\frac{M_{DM}}{M_\oplus}\right)^{5/4} \approx 2.\!\times\!10^4 \left(\frac{M_*}{M_\oplus}\right), \qquad (10)$$

Since in our model $M_*$ is unconstrained Eq. 10 formally allows the transition from a dark energy star dominated universe to a radiation dominated universe to take place for any value for $M_{DM}$. However this transition cannot occur so late that it interferes with the requirement that the cosmological production of helium and other light elements should be approximately the same as in the standard cosmological model. This constraint limits $1+z_r$ to be $>10^{10}$ and $M_{DM} < 2\times10^4 M_\oplus$ One may also invoke the limits on the present day abundance of MACHO objects set by gravitational micro-lensing [16] to say that $M_{DM}$ should be $>10 M_\oplus$. In the following we will adopt as our *a priori* range for the average primordial compact object mass $2\times10^4 M_\oplus > M_{DM} > 10 M_\oplus$. For these nominal values of the dark matter masses the CMB originates at a red shift in the range $5\times10^{11} > 1+z_r > 10^{10}$. The radiation temperature at redshift $z_r$ would lie in the range $120 MeV > T(z_r) > 2.6 MeV$, which for the most part is above the temperatures where cosmological production of the light elements takes place.

Eq. 10 implies that for our assumed range of dark matter masses the mass of the initial primordial dark energy stars lies in the range $12 M_\oplus > M_* > 9\times10^{-4} M_\oplus$. The initial positive vacuum energy density $\rho_*$ is related to $M_*$ by $\rho_* = 0.1 GeV^4 (M_\oplus/M_*)^2$, which just expresses the fact that for a dark energy star $M$ is the mass of the vacuum energy inside the de Sitter horizon. The limits on $M_*$ derived from Eq. 10 translate to $10^5 GeV^4 > \rho_* > 7\times10^{-4} GeV^4$. Given the present day cosmological density of dark matter ($2\times10^{-30}$ gm/cc) the redshift where the dark energy stars were initially formed can be found from Eq, 6, and lie in the range $5\times10^{11} > 1+z_* > 10^{10}$ for our nominal range for $M_{DM}$. By construction the ranges for $M_*$ and $\rho_*$ just quoted are consistent with the present day density of dark matter. However Eq. 8 also yields a present day radiation temperature that is very close to the observed of the CMB temperature for all values of $M_{DM}$ in our nominal range.

A very encouraging prediction of our model follows from the fact that the initial metric fluctuations created by the quantum instability of de Sitter horizons have the Harrison-Zeldovich-Peebles form: [17-19]:

$$\frac{\delta\rho}{\rho} \approx \varepsilon_0 (R_0 k)^2 \qquad (11)$$

where $R_0 = 2GM^*/c^2$ is the gravitational radius corresponding to the initial positive vacuum energy, $\varepsilon_0 \sim 1$ is the metric fluctuation created on the scale $R_0$ by the formation the objects with mass $M^*$, and $\delta\rho/\rho$ is the fractional density fluctuation for scales $k^{-1} >> R_0$. Because the speed of sound in an the expanding universe of dark energy stars is very low, the density fluctuations will rapidly grow until the radiation locked up as the energy of excited dark energy star becomes freely streaming. According to the Lifschitz formula [20] for the growth of density fluctuations during a matter dominated period, by the time the redshift decreases to $1+z_r$, the fluctuations in the spatial density of primordial dark energy stars will have grown by a factor $(1+z_*/1+z_r)$, independent of length scale. Taking this into account, and averaging the density fluctuations predicted by Eq. 11 over all volumes that could have collapsed by the time that the expanding universe had reached the beginning of the radiation dominated era at redshift $1+z_r$, we obtain [see Ref9 for details] as an estimate for the renormalized value of $\varepsilon_0$ at redshift $1+z_r$ :

$$\varepsilon_r \approx 3\varepsilon_0 \left[\frac{1+z_r}{1+z_*}\right]^2 . \qquad (12)$$

For our assumed range of dark matter average masses our model for the big bang predicts that

$$1.2 \times 10^{-5} > (1+z_r/1+z_*)^2 > 10^{-6} \qquad (13)$$

Considering the simplicity of our model, these values are in remarkably good agreement with the observed value, $\delta T/T \sim 10^{-5}$, for the mean temperature fluctuation of the CMB, which corresponds to $\varepsilon_r \sim 3 \times 10^{-5}$. Taken literally Eq. 12 suggests that the average dark mass compact object has a mass close to $10^4$ solar masses.

It is of course a dramatic prediction of our model suggests that dark matter consists of compact objects with masses on the order of

$10^4 M_\oplus$. Actually it is an old idea that dark matter consists of primordial black holes (PBHs) [21-24], although at the time this was first proposed there was no preference for the typical masses of the PBHs. The possibility that dark matter may consist of "lumps of false vacuums" has also been considered [25]. Recently the idea that dark matter consists of compact objects with many solar masses has received renewed interest as a result of the failure to identify any stable elementary particle that might serve as a candidate for dark matter [26,27]. As it happens there are at the present time no astronomical observations which can definitively rule out this possibility if the masses of the dark matter compact objects are $>10 M_\oplus$. It should be noted in this connection that because the dark energy star critical frequency $h\nu_c > 7MeV$ for the range of dark matter object masses we are considering, our dark matter compact objects will be transparent at x-ray wavelengths. Therefore Bondi-Hoyle accretion onto our dark matter objects will not produce any significant x-ray emission. However there may be significant $\gamma$-ray emissions [28].

As noted earlier gravitational microlensing searches for dark compact objects have ruled out the rule out the possibility that dark matter consists primarily of compact objects with masses < $10 M_\oplus$ [17]. Unfortunately because the gravitational lens brightening of a background star due to a compact object with a mass greater than $100 M_\oplus$ would last for years, it would be difficult to use the micro-lensing techniques previously used to identify Macho objects like those contemplated in this paper. However, newer types of instruments, such as the LSST, may make it possible to identify intermediate mass dark matter objects [T. Axelrod, private communication.].1

Evidently the all the features of the CMB as well as many features of dark matter follow from our hypothesis that the big bang created a positive vacuum energy with an energy density > $(GeV)^4$. Rather amusingly our predictions for the nature of dark matter are *ipso facto* completely consistent with the observed inhomogeneity of matter at practically all scales. Indeed state of the art numerical simulations of the evolution of dark matter structures use point particles with a fixed mass typically in the range $10^3 – 10^4 M_\oplus$ [for a review see 29]. Furthermore in order to simulate of the formation of galactic structures within the framework of the numerical models for the evolution of dark matter structures it is necessary to add primordial point-like seed masses of about $10^5$ solar masses in order to obtain the observed galactic morphologies [30]. Of course it follows from our prediction that the dark matter compact objects were formed from the stochastic coalescence of primordial dark energy stars with mass $M_*$ that compact objects with masses larger than the average mass $M_{DM}$ were also formed. In conclusion we note that our model for the big bang is the only cosmological model known to us that provides a simple and direct explanation for the large scale features of our universe.

## Acknowledgements
The authors are very grateful to T. Axelrod, P. Frampton, C. Frenk, and A. Loeb for discussions regarding observational constraints on many solar mass dark matter constituents.